\documentclass[12pt]{article}
\usepackage[dvips]{color}
\usepackage{epsfig}
\usepackage{amsmath}
\usepackage{graphicx}

\begin{document}
\title{Strong Gravitational Lensing in a Charged Squashed Kaluza- Klein G\"{o}del Black hole}
\author{{J. Sadeghi $^{a}$\thanks{Email: pouriya@ipm.ir}\hspace{1mm}
and {H. Vaez $^{a}$\thanks{Email: h.vaez@umz.ac.ir}\hspace{1mm}
}}\\
{$^{a}$ \emph{Physics Department, Mazandaran University},}\\{
\emph{P.O.Box 47416-95447, Babolsar, Iran}}\\} \maketitle
\begin{abstract}
In this paper we investigate the strong gravitational lansing in a
charged squashed Kaluza-Klein G\"{o}del black hole. The deflection
angle is considered by the logarithmic term proposed by Bozza et al.
Then we study the variation of deflection angle and its parameters
$\bar{a}$ and $\bar{b}$ . We suppose  that the supermassive black
hole in the galaxy center can be considered by a charged squashed
Kaluza-Klein black hole in a G\"{o}del background and by relation
between lensing parameters and observables, we estimate the
observables for different values of charge, extra dimension and
G\"{o}del parameters. \noindent
\end{abstract}
\section{Introduction}
As we know  the light rays or photons would be deviated from their
straight way when they pass close to the massive object such as
black holes. This deflection of light rays is known as gravitational
lensing. This gravitational lensing is one of the applications and
results of general relativity \cite{Einstein} and is used as an
instrument in astrophysics, because it can  help us to extract the
information about stars. In 1924 Chwolson pointed out that when a
star(source), a deflector(lens) and an observer are perfectly
aligned, a ring-shape image of the star appears which is called
'Einstein ring'. Other studies have been led by  Klimov, Liebes,
refsdal and Bourassa and Kantowski \cite{Bourassa}. Klimov
investigated the lensing of galaxies by galaxies \cite{klimov}, but
Liebes
 studied the lensing of stars by stars and also stars
by clusters in our galaxy \cite{Liebes}. Refsdal showed that the
geometrical optics can be used for investigating the gravitational
lenses properties and time delay resulting from it
\cite{Refsdal1,Refsdal2}.
 The gravitational lensing
has been presented in details  in \cite{schneider} and  reviewed  by
some papers (see for examples \cite{Narayan}-\cite{wabsganss}). At
this stage, the gravitational lensing is developed for weak field
limit and could not describe some phenomena such as looping of light
rays near the massive objects. Hence, scientists started to study
these phenomena from another point of view and they proposed
gravitational lensing in a strong field limit. When the source is
highly aligned with lens and the observer, one set of infinitive
relativistic "ghost" images would be produce on each side of black
hole. These images are produced when the light rays that pass very
close to black hole, wind one or several times around the black hole
before reaching to observer. At first, this phenomenon was proposed
by Darwin \cite{Darwin} and revived in Refs.
\cite{Chandrasekhar}-\cite{luminet}. Darwin
 proposed a surprisingly easy formula for the positions of
the relativistic images generated by a Schwarzschild black hole.
Afterward several studies of null geodesics in strong gravitational
fields have been led in literatures: a semi-analytical investigation
about geodesics in Kerr geometry has been made in \cite{Viergutz},
also the appearance of a black hole in front of a uniform background
was studied in Refs. \cite{Bardeen,Falcke}. Recently, Virbhadra and
Ellis formulated lensing in the "strong field limit" and obtain the
position and magnification of these images for the Schwarzschild
black hole \cite{Virbhadra1,Virbhadra2}. In Ref \cite{Frittelli}, by
an alternative formulation,  Frittelli, Kiling and Newman attained
an exact lens equation, integral expressions for its solutions, and
compared their result with Virbhadra and Ellis. Afterwards, the new
method was proposed by Bozza et al. in which they revisited the
schwarzschild black hole lensing by retaining the first two leading
order terms \cite{bozza2}. This technic was used by Eiroa, Romero
and Torres to study a Reissner-Nordstrom black hole \cite{Eiroa} and
Petters to calculate relativistic effects on microlensing events
\cite{petters}. Finally, the generalization of Bozza's method for
spherically symetric metric was developed in \cite{bozza1}. Bozza
compared the image patterns for several interesting backgrounds and
showed that by the separation of the first two relativistic images
we can distinguish two different collapsed objects. Further
 development for other black holes can be found in \cite{bozza5}-\cite{saadat}. Several interesting
 studeies are devoted to lensing by  naked singularities \cite{naked,Virbhadra3},
Janis-Newman-Winicour metric  \cite{bozza1} and role of scaler field
in gravitational lensing
 \cite{Virb}. In Ref \cite{Virb},
Virbhadra  et al. have considered a static and circularly symmetric
lens characterized by
mass and scalar charge parameter and investigated the lensing for different values of charge parameter.\\
The gravitational lenses are important tools for probing the
universe. Narasimha and Chitre predicted that the gravitational
lening of dark matter can give the useful data about of position of
dark matter in the universe \cite{NarasimhaChitre,massey}. Also in
some papers gravitational lens is used to detect exotic objects in
the universe,
 such as cosmic strings \cite{hogan}-\cite{Vilenkin2}. \\
Recently, the idea of large extra dimensions has attracted much
attention to construct theories in which gravity is unified with
other forces \cite{Emparan}. One of the most interesting problems is
the verification of extra dimensions by physical phenomena. For this
purpose  higherdimensional black holes in accelerators
\cite{higher1,higher2} and in cosmic rays
\cite{Arkani}-\cite{Anchordoqui} and gravitational waves from
higher-dimensional black holes \cite{Seahra} are studied. The
five-dimensional Einstein-Maxwell theory with a Chern- Simons term
predicted five-dimensional charged black holes \cite{Gunaydin}. Such
a higher-dimensional black holes would reside in a spacetime that is
approximately isotropic in the vicinity of the black holes, but
effectively four-dimensional far from the black holes
\cite{fourdim}. These higher dimensional black holes are called
Kaluza-Klein black holes. The presence of extra dimension is tested
by quasinormal modes from the perturbation around the higher
dimensional black hole \cite{QNM1}-\cite{QNM7} and the spectrum of
Hawking radiation \cite{Hawkingradiation1}-\cite{Hawkingradiation4}.
The gravitational lensing is another method to investigate the extra
dimension. Thus, the study of strong gravitational lensing by higher
dimensional black hole can help us to extract  information about the
extra dimension in astronomical observations in  the future. The
Kaluza-Klein black holes with squashed horizon \cite{ishihara} is
one of the extra dimensional black holes and it's Hawking radiation
 and quasinormal modes have been
investigated in some papers
\cite{HawkingradiationSKKBH1}-\cite{QNMSKKBH2}. Also, the
gravitational lensing of these black holes is studied in several
papers. Liu et al. have studied the gravitational lensing by
squashed Kaluza-Klein black holes  in Refs \cite{SKKBH,SKKGBH} and
Sadeghi et al. investigated the charged type of this black hole
\cite{CSKKBH}. \\
One the other hands we know that our universe is rotational and it
is reasonable to consider G\"{o}del background for our universe. An
exact solution for rotative universe was obtained by G\"{o}del. He
solved Einstein equation with pressureless matter and negative
cosmological constant \cite{Godel}. The solutions representing the
generalization of the G\"{o}del universe in the minimal five
dimensional gauged supergravity are considered in many studies
\cite{Wu}-\cite{Kunduri}.  The properties of various black holes in
the G\"{o}del background are investigated in many works
\cite{gghghghhg}. The strong gravitational lensing in a Squashed
Kaluza-Klein Black hole in a G\"{o}del universe  is
investigated in Ref. \cite{SKKGBH}. \\
In this paper, we tudy the strong gravitational lensing in a charged
squashed Kaluza-Klein G\"{o}del black hole. In that case, we see the
effects of the scale of the extra dimension, charge of black hole
and G\"{o}del parameter on the coefficients and observables of
strong gravitational lensing.\\
So, this paper is organized as follows: Section 2 is briefly devoted
to charged squashed Kaluza-Klein G\"{o}del black hole background. In
section 3 we use the Bozza's method \cite{bozza5,bozza3} to obtain
the deflection angle and other parameters of strong gravitational
lensing as well as  variation of them with extra dimension,
G\"{o}del parameter and charge of black hole . In section 4, we
suppose that the supermassive object at the center of our galaxy can
be considered by the metric of charged squashed Kaluza-Klein
G\"{o}del black hole. Then, we evaluate the numerical results for
the coefficients and observables in the strong gravitational lensing
. In the last Section, we present a summary of our work.
\section{The charged squashed Kaluza- Klein G\"{o}del black hole metric}
The charged squashed Kaluza- Klein G\"{o}del black hole spacetime is
given by \cite{Wu},
\begin{equation}\label{metric1}
ds^2=-f(r)dt^2+\frac{k^2(r)}{V(r)}dr^2-2g(r)\sigma_3
dt+h(r)\sigma^2_3+\frac{r^2}{4}[k(r)(\sigma^2_1+\sigma^2_2)+\sigma^2_3],
\end{equation}
where
\begin{eqnarray}\label{sigma}
&&\sigma_1=\cos\psi\, d\theta+\sin\psi\, \sin\theta\, d\phi,\nonumber\\
&&\sigma_2=-\sin\psi\,d\theta+\cos\psi\,\sin\theta\,d\phi,\nonumber\\
&&\sigma_3=d\psi+\cos\theta\, d\phi.
\end{eqnarray}
\begin{eqnarray}\label{function}
f(r)=1-\frac{2M}{r^2}+\frac{q^2}{r^4},\,\,\,\,\,\,\,g(r)=j(r^2+3q),\,\,\,\,\,h(r)=-j^2
r^2(r^2+2M+6q),\,\,\,\,\,\,\nonumber\\
V(r)=1-\frac{2M}{r^2}+\frac{16j^2(M+q)(M+2q)}{r^2}+\frac{q^2(1-8j^2(M+3q))}{r^4},\,\,\,\,\,\,
k(r)=\frac{V(r_\infty)r_\infty^4}{(r^2-r_\infty^2)^2}.
\end{eqnarray}
and $0\leq\theta<\pi$, $0\leq\phi<2\pi$, $0\leq\psi<4\pi$ and
$0<r<r_\infty$. Here $M$ and $q$ are the mass and charge of the
black hole respectively and $j$ is the parameter of G\"{o}del
background. The killing horizon of the black hole is given by
equation
 $V(r)=0$ , where
\begin{eqnarray}\label{horizon1}
r_{h}^2=M-8j^2(M+q)(M+2q)\pm\sqrt{[M+q-8j^2(M+2q)^2][M-q-8j^2(M+q)^2]}.
\end{eqnarray}\\
We see that the black hole has two horizons. As $q\longrightarrow 0$
the horizon of the squashed Kaluza- Klein G\"{o}del black hole is
recovered \cite{SKKGBH} and when $q$ and $j$ tend to zero, we have
$r_{h}^2=2M$, which is the horizon of five-dimensional Schwarzschild
black hole. Here we note that the argument of square root
constraints the mass, charge and G\"{o}del parameter values. When
$r_{\infty}\longrightarrow \infty$, we have $k(r)\longrightarrow1$,
which means that the squashing effect
disappears and the five-dimensional charged black hole is recovered.\\
By using the transformations,
$\rho=\rho_0\frac{r^2}{r^2_\infty-r^2}$,
$\tau=\sqrt{\frac{\rho_0(1+\alpha)}{\rho_0+\rho_M}}t$ and
$\alpha=\frac{\rho_q^2(\rho_0+\rho_M)}{\rho_0(\rho_0+\rho_q)^2}$,
the metric ~(\ref{metric1}) can be written in the following form,
\begin{equation}\label{metric2}
ds^2=-\mathcal{F}(\rho)d\tau^2+\frac{K(\rho)}{\mathcal{G}(\rho)}d\rho^2+\mathcal{C}(\rho)(d\theta^2+sin^2\theta\,d\phi^2)-2H(\rho)\sigma_3d\tau+\mathcal{D}(\rho)\sigma_3^2,
\end{equation}
\begin{eqnarray}\label{function2}
&&\mathcal{F}(\rho)=1-\frac{\rho_M-2\alpha\rho_0}{(1+\alpha)\rho_M}(\frac{\rho_{M}}{\rho})+
\frac{\alpha}{1+\alpha}(\frac{\rho_0}{\rho_M})^2(\frac{\rho_M}{\rho})^2,\nonumber\\
&&K(\rho)=1+\frac{\rho_0}{\rho}\,,\,\,\,\,\,\mathcal{G}=(1-\frac{\rho_{h+}}{\rho})(1-\frac{\rho_{h-}}{\rho}),\nonumber\\
&&\mathcal{C}(\rho)=\rho^2
K(\rho),\,\,\,\,H(\rho)=jr_\infty^2\left(\frac{1}{K(\rho)}+\frac{3\rho_q}{\rho_0+\rho_q}\right)\sqrt{\frac{\rho_0+\rho_M}{\rho_0(1+\alpha)}},\nonumber\\
&&\mathcal{D}(\rho)=\frac{r^2_\infty}{4K(\rho)}-\frac{j^2\rho
r_\infty^2 }{(\rho+\rho_0)^2(\rho_M+\rho_0)(\rho_q+\rho_0)}\times\nonumber\\
&&\left\{\rho[\rho_0(\rho_0+2\rho_M)+7\rho_0\rho_q+8\rho_M\rho_q]+\rho_0[\rho_0(\rho_M+6\rho_q)+7\rho_M\rho_q]\right\},
\end{eqnarray}
with
\begin{eqnarray}\label{tranformations}
\rho_M=\rho_0\frac{2M}{r^2_\infty-2M}\,\,,
\,\,\,\,\,\,\,\rho_q=\rho_0\frac{q}{r^2_\infty-q},\,\,\,\,\,\,\rho_{h\pm}=\rho_0\frac{r_{h\pm}^2}{r^2_\infty-r_{h\pm}^2}\,.
\end{eqnarray}
\epsfxsize=18cm \epsfysize=20cm
\begin{figure}
\centerline{\epsffile{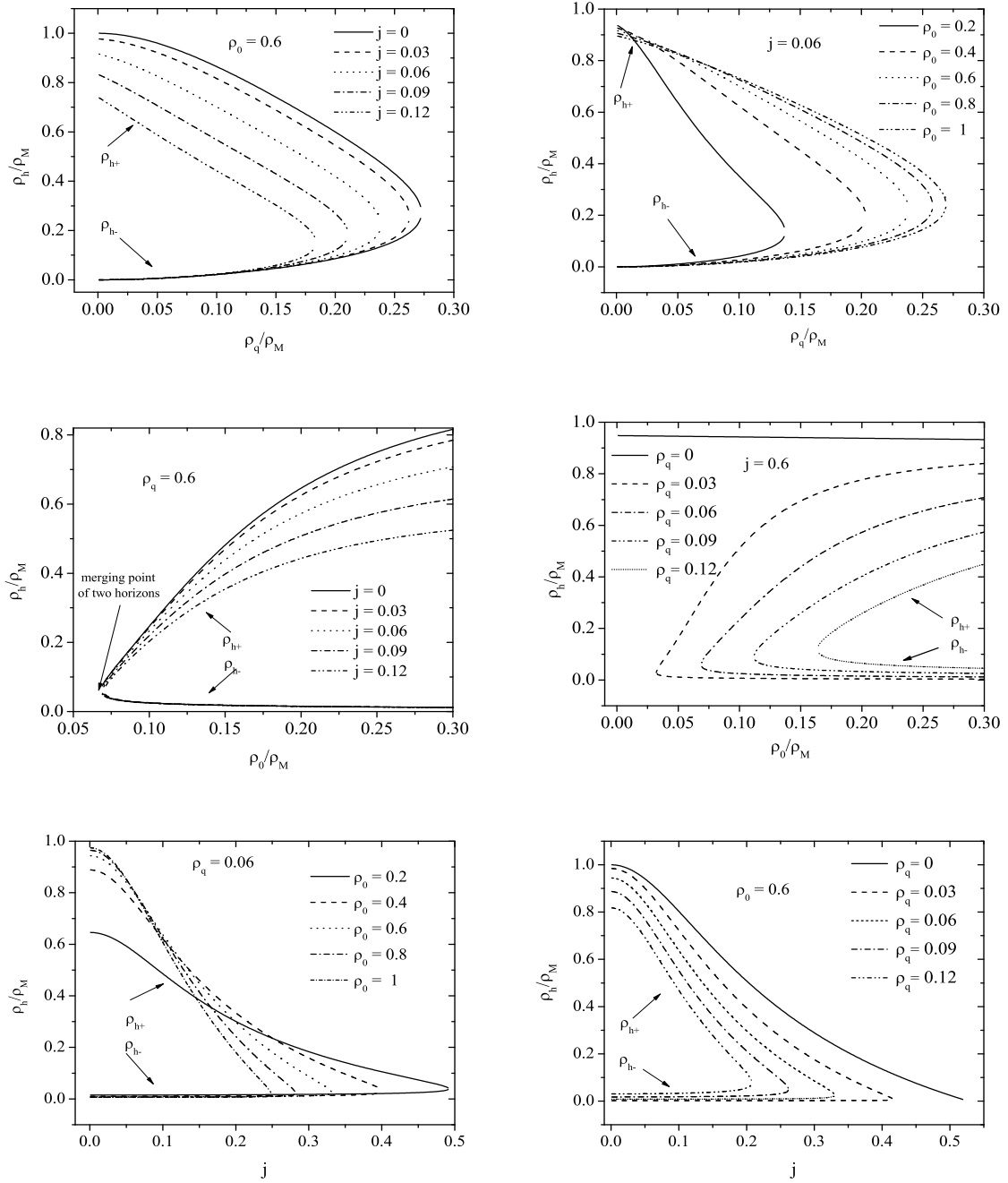}} \caption{The plots show the
variation  of horizon radiuses with respect to $j$, $\rho_0$ and
$\rho_q$ (Note that in each figure, for $\rho_q\neq0$, two horizons
merge
 at a point. This point has been shown for one of figures. )}
\end{figure}
Where $\rho_{h+}$ and $\rho_{h-}$ denote the outer and inner
horizons of the black hole in the new coordinate and $\rho_0$ is a
scale of transition from five-dimensional spacetime to an effective
four-dimensional one. Here
$\rho_0^2=\frac{r^2_\infty}{4}V(r_\infty)$, so that
$r^2_\infty=4(\rho_0+\rho_{h+})(\rho_0+\rho_{h-})$. The Komar mass
of black hole is related to $\rho_M$ with $\rho_M=2G_4M$, where
$G_4$ is the four dimensional  gravitational constant. By using
relations~(\ref{horizon1}) and~(\ref{tranformations}) we can obtain
$\rho_{h\pm}$ in the following coupled equations,
\begin{eqnarray}\label{equations}
&&2\left[\rho_0(\rho_{h+}+\rho_{h-})+2\rho_{h+}\rho_{h-}\right]={a}(\rho_{h+},\rho_{h-}),\nonumber\\
&&2\left[\rho_0(\rho_{h+}-\rho_{h-})\right]={b}(\rho_{h+},\rho_{h-}),
\end{eqnarray}
where
\begin{eqnarray}\label{ab}
&&a=\frac{\rho_M
r_\infty^2}{2(\rho_0+\rho_M)}-2j^2r_\infty^4\frac{\left(\rho_M\rho_0+3\rho_M\rho_0+2\rho_q\rho_0\right)\left(\rho_M\rho_0+5\rho_M\rho_q+4\rho_q\rho_0\right)}{(\rho_0+\rho_M)^2(\rho_0+\rho_q)^2},\nonumber\\
&&b=\left\{a^2-4\frac{\rho_q^2r_\infty^4}{(\rho_0+\rho_q)^2}\left(1-4j^2r_\infty^2\frac{(\rho_0\rho_M+7\rho_M\rho_q+6\rho_q\rho_0)}{(\rho_0+\rho_M)(\rho_0+\rho_q)}\right)\right\}^{\frac{1}{2}}.
\end{eqnarray}
when $\rho_q\longrightarrow0$, the horizon of black hole becomes
$\rho_h=\frac{2(\rho_0+\rho_M)}{\sqrt{1+64j^2\rho_M^2}+}-\rho_0$, as
obtained in \cite{SKKGBH}. In case of  $\rho_q\longrightarrow0$ and
$j\longrightarrow0$, we have $\rho_h=\rho_M$ which is consistent
with neutral squashed Kaluza-Klein black hole \cite{SKKBH}. You note
that the square root in relation ~(\ref{ab}) constrains the values
of $\rho_0$, $\rho_q$ and $j$. In the case $j=0$, the permissive
regime  is shown in Ref.\cite{CSKKBH}. For any value of $\rho_q$
there is allowed rang for $\rho_0$. Hence these parameters can not
select any value and when $j$ increases from zero, permissive regime
for $\rho_q$ and $\rho_0$ becomes more confined. We solved the above
coupled equations numerically and results are shown in figure 1. We
see that the outer horizon of black hole increases with the size of
extra dimension, $\rho_0$ and decreases with $j$ and $\rho_q$. Note
that two horizons of black hole coincide in especial values of
parameters, which in that case we have an extremal black hole.
\section{Geodesic equations, Deflection angle}
In this section, we are going to investigate the deflection angle of
light rays when they pass close to a charged squashed Kaluza-Klein
G\"{o}del black hole. We also study the effect of the charge
parameter $\rho_q$, the scale of extra dimension $\rho_0$ and
G\"{o}del parameter on the deflection angle and it's coefficients in
the equatorial plane
($\theta=\pi/2$).\\
In this plane, the squashed Kaluza-Klein G\"{o}del metric reduces to
\begin{equation}\label{metric3}
ds^2=-\mathcal{F}(\rho)d\tau^2+\mathcal{B}(\rho)d\rho^2+\mathcal{C}(\rho)\,d\phi^2+\mathcal{D}(\rho)d\psi^2-2H(\rho)dt
d\psi,
\end{equation}
where
\begin{equation}\label{B}
\mathcal{B}(\rho)=\frac{K(\rho)}{\mathcal{G}(\rho)}.
\end{equation}
The null geodesic equations are,
\begin{equation}\label{geodesic}
{\ddot{x}_i}+\Gamma_{jk}^i\,\dot{x}^j\,\dot{x}^k=0,
\end{equation}
where
\begin{equation}\label{geodesic2}
g_{ij}\dot{x}^i\,\dot{x}^j=0,
\end{equation}
where $\dot{x}$ is the tangent vector to the null geodesics and the
dote denotes derivative with respect to affine parameter. We use
equation (\ref{geodesic}) and obtain the following equations,
\begin{eqnarray}\label{constMo}
&&\dot{t}=\frac{\mathcal{D}(\rho)E-H(\rho)L_\psi}{H^2(\rho)+\mathcal{F}(\rho)\mathcal{D}(\rho)},\nonumber\\
&&\dot{\phi}=\frac{L_\phi}{\mathcal{C}(\rho)},\nonumber\\
&&\dot{\psi}=\frac{H(\rho)E+\mathcal{F}(\rho)L_\psi}{H^2(\rho)+\mathcal{F}(\rho)\mathcal{D}(\rho)},
\end{eqnarray}
\begin{equation}\label{rhoGeo}
(\dot{\rho})^2=\frac{1}{\mathcal{B}(\rho)}\left[\frac{\mathcal{D}(\rho)E-2H(\rho)EL_\psi-\mathcal{F}(\rho)L^2_\psi}{H^2(\rho)
+\mathcal{F}(\rho)\mathcal{D}(\rho)}-\frac{L^2_\phi}{\mathcal{C}(\rho)}\right].
\end{equation}
where $E$, $L_\phi$ and $L_\psi$ are constants of motion. Also, the
$\theta$-component of equation~(\ref{geodesic}) in equatorial plane
$\theta=\pi/2$, is given by,
\begin{eqnarray}\label{s14}
\dot{\phi}\left[\mathcal{D}(\rho)\dot{\psi}-H(\rho)\dot{t}\right]=0.
\end{eqnarray}
If $\dot{\phi}=0$, then deflection angle will be zero and this is
illegal, So we set
$L_\psi=\mathcal{D}(\rho)\dot{\psi}-H(\rho)\dot{t}=0$. By using
equation~(\ref{rhoGeo}) one can obtain following expression for the
impact parameter,
\begin{equation}\label{impact}
L_{\phi}=u=\sqrt{\frac{\mathcal{C}(\rho_s)\mathcal{D}(\rho_s)}{H^2(\rho_s)+\mathcal{F}(\rho_s)\mathcal{D}(\rho_s)}},
\end{equation}
and the minimum of impact parameter takes place in photon sphere
radius $r_{ps}$, that is given by the root of following equation
\cite{photonsphere},
\begin{equation}\label{phEqu}
\mathcal{D}(\rho_s)\,\left[H(\rho_s)^2+\mathcal{F}(\rho_s)\mathcal{D}(\rho_s)\right]\mathcal{C}^\prime(\rho_s)-
\mathcal{C}(\rho_s)\,\left[\mathcal{D}(\rho_s)^2\mathcal{F}^\prime(\rho_s)+
2\mathcal{D}(\rho_s)H(\rho_s)H^\prime(\rho_s)-H(\rho_s)^2\mathcal{D}^\prime(\rho_s)\right]=0.
\end{equation}
\begin{figure}
\centerline{\epsffile{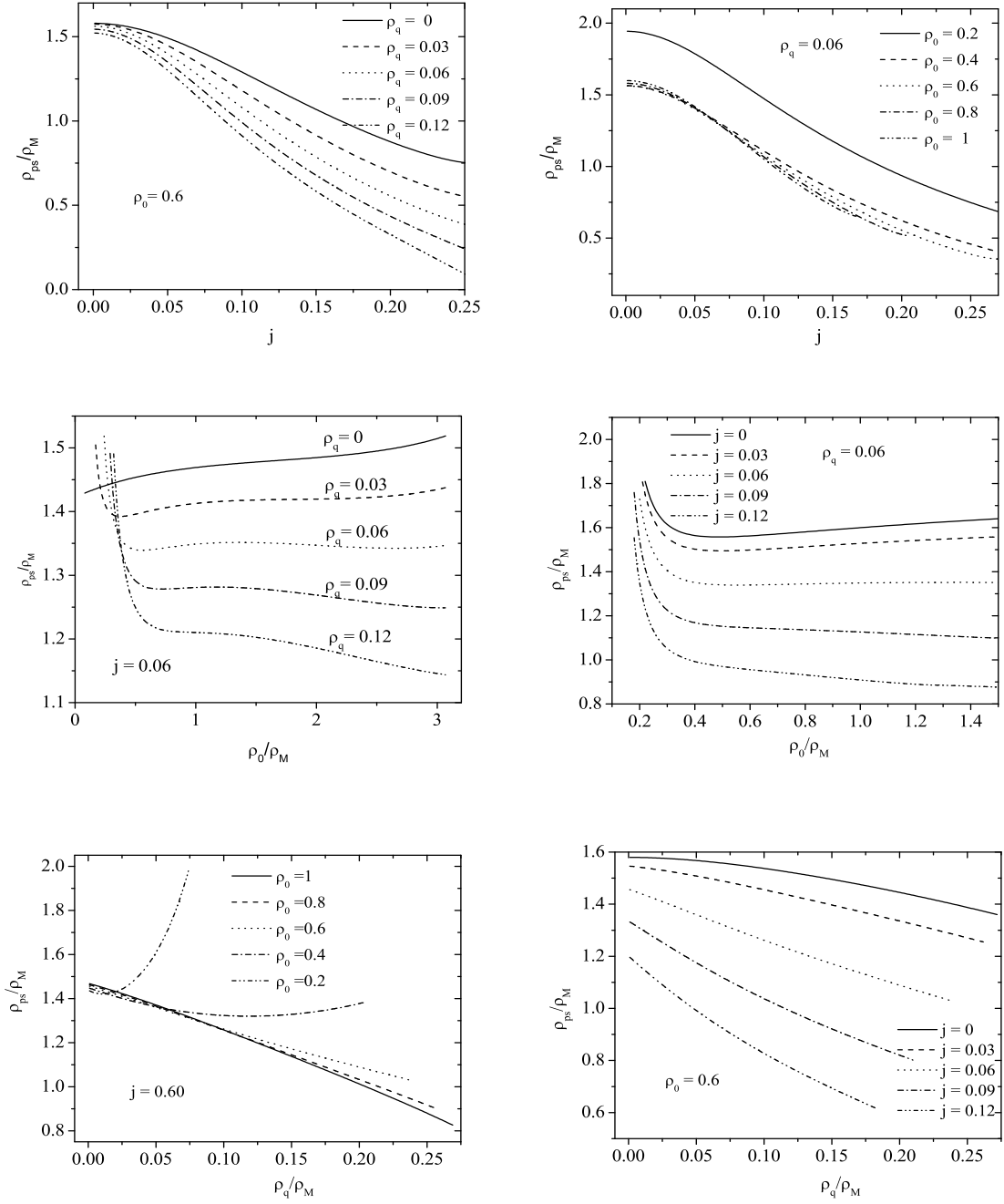}} \caption{The
 variation of photon sphere radius with respect to $j$, $\rho_0$ and $\rho_q$.}
\end{figure}

\begin{figure}
\centerline{\epsffile{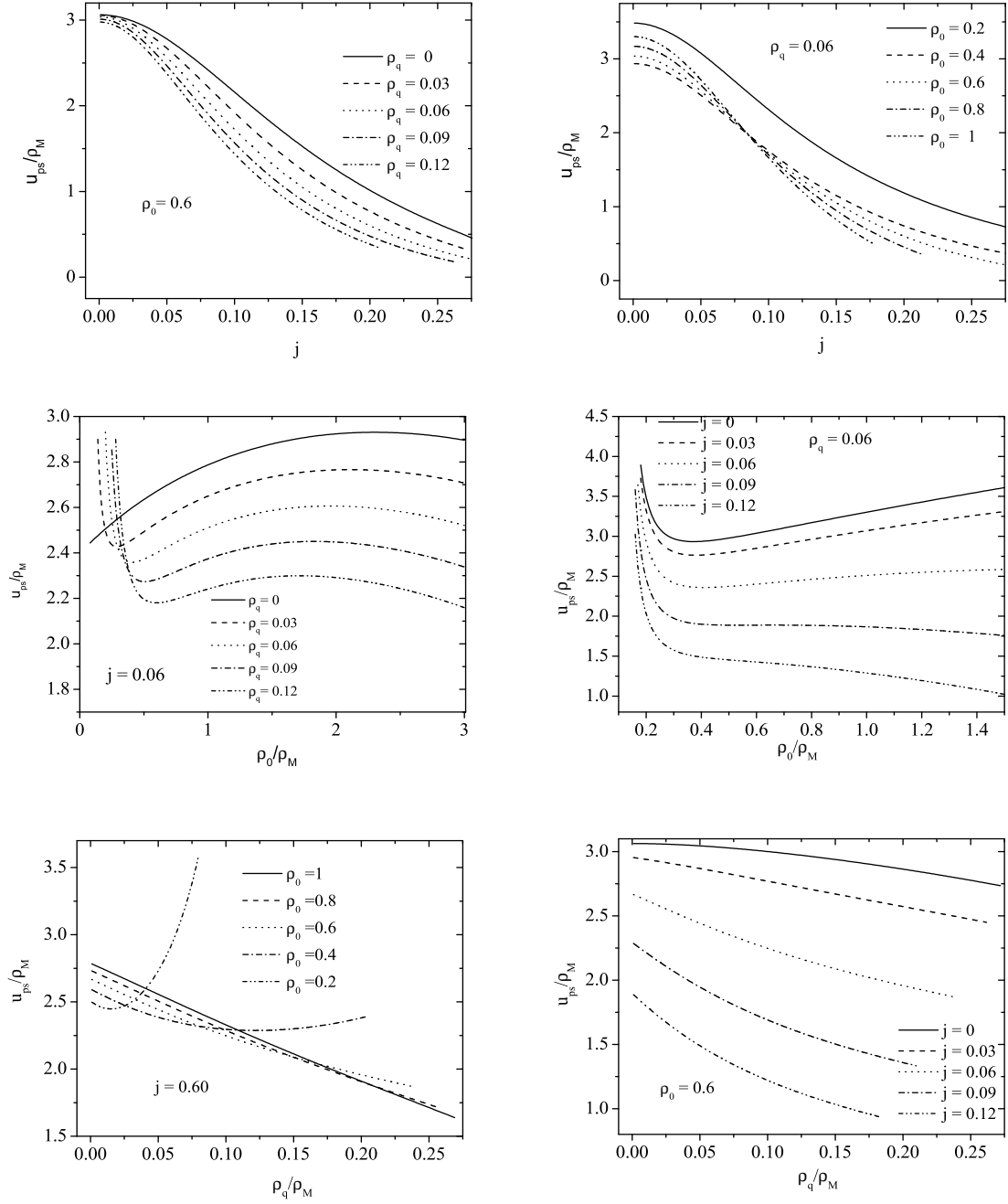}} \caption{The
 variation  of impact parameter as a function of $j$, $\rho_0$ and $\rho_q$.}
\end{figure}

\begin{figure}
\centerline{\epsffile{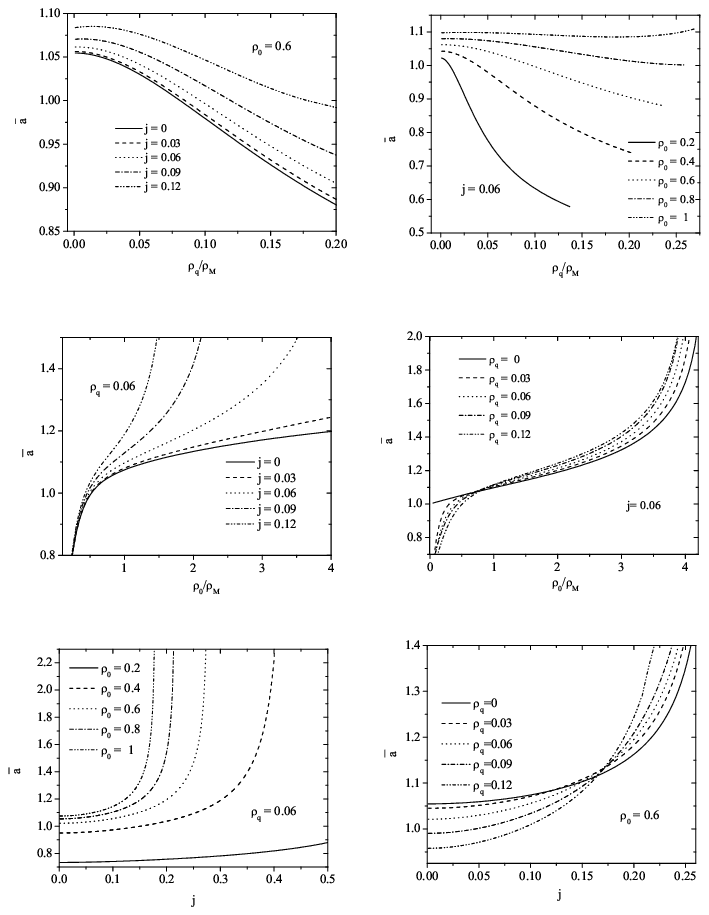}} \caption{The
 variation  of $\bar{a}$ with respect to $j$, $\rho_0$ and $\rho_q$.}
\end{figure}
Here $\rho_s$ is the closet approach for light ray and the prime is
derivative with respect to $\rho_s$. The analytical  solution for
the above equation is very complicated, so we calculated the
equation (\ref{phEqu}) numerically. Variations of r photon sphere
radius are plotted with respect to the charge $\rho_q$, the scale of
extra dimension $\rho_0$ and G\"{o}del parameter in the figure 2.
Also figure 3 shows variations of impact parameter in it's minimum
value (at radius of photon sphere). These figures show that by
adding the charge to the black hole, the behavior of the photon
sphere radius
 and minimum of impact parameter is different
compare with the neutral black hole \cite{SKKBH}. As $\rho_0$
approaches to it's minimum values the radius of photon sphere and
impact parameter become divergent.\\
By using the chain derivative and equation (\ref{rhoGeo}), the
deflection angle in the charged squashed Kaluza-Klein G\"{o}del
black hole can be written as,
\begin{eqnarray}\label{s17}
\alpha_\varphi(\rho_s)=I_\varphi(\rho_s)-\pi,\nonumber\\
\alpha_\psi(\rho_s)=I_\psi(\rho_s)-\pi,
\end{eqnarray}
\begin{eqnarray}\label{Iphi}
I_\varphi(\rho_s)=2\int^\infty_{\rho_s}\frac{\sqrt{\mathcal{B}(\rho)\mathcal{A}(\rho)\mathcal{C}(\rho_s)}}{\mathcal{C}(\rho)}
\frac{1}{\sqrt{\mathcal{F}(\rho_s)-\mathcal{F}(\rho)\frac{\mathcal{C}(\rho_s)}{\mathcal{C}(\rho)}}}\,\,\,d\rho,
\end{eqnarray}
\begin{eqnarray}\label{Ipsi}
I_\psi(\rho_s)=2\int^\infty_{\rho_s}\frac{H(\rho)}{\mathcal{D}(\rho)}\sqrt{\frac{\mathcal{B}(\rho)\mathcal{A}(\rho_s)}{\mathcal{A}(\rho)}}
\frac{1}{\sqrt{\mathcal{F}(\rho_s)-\mathcal{F}(\rho)\frac{\mathcal{C}(\rho_s)}{\mathcal{C}(\rho)}}}\,\,\,d\rho,
\end{eqnarray}
with
\begin{equation}\label{A}
\mathcal{A}(\rho)=\frac{H^2(\rho)+\mathcal{F}(\rho)\mathcal{D}(\rho)}{\mathcal{D}(\rho)}.
\end{equation}
 When we decrease the $\rho_s$ (and consequently $u$) the
deflection angle increases. At some points, the deflection angle
 exceeds  from $2\pi$ so that the light ray will make a complete loop
around the compact object before reaching at the observer. By
decreasing $\rho_s$ further, the photon will wind several times
around the black hole before emerging. Finally, for
$\rho_s=\rho_{sp}$ the deflection angle diverges and the photon is
captured by the black hole. Moreover, from equations (\ref{Iphi})
and (\ref{Ipsi}), we can find that in the Charged Squashed
Kaluza-Klein G\"{o}del black hole, both of the deflection angles
depend on the parameters $j$, $\rho_0$ and $\rho_q$, which implies
that we could detect the rotation of universe, the extra dimension
and charge of black hole in theory by gravitational lens. Note that
the $I_\phi(\rho_s)$ depend on $j^2$, not $j$. It
 shows that the deflection angle is independent of the direction of rotation of universe.
  But, from equation (\ref{Ipsi}) we find that the integral
  $I_\psi(\rho)$ contains the factor $j$, then the deflection angle
  $\alpha_\psi(\rho_s)$ for the photon traveling around the lens
  in two opposite directions  is different .The main
  reason is that the equatorial plan is parallel with G\"{o}del
  rotation plan \cite{SKKGBH}. When $j$ vanishes, the deflection
  angle of $\psi$ tends zero \cite{SKKBH, CSKKBH}. We  focus on the deflection angle in the $\phi$ direction,
  \\
   So we can rewrite the equation~(\ref{Iphi}) as,
\begin{equation}\label{Iz}
I(\rho_s)=\int^1_0R(z,\rho_s)f(z,\rho_s)\,dz,
\end{equation}
with
\begin{equation}\label{R}
R(z,\rho_s)=2\frac{\rho}{\rho_{s}\mathcal{C}(\rho)}\sqrt{\mathcal{B}(\rho_s)\mathcal{A}(\rho)\mathcal{C}(\rho_s)},
\end{equation}
and
\begin{equation}\label{f(z)}
f(z,\rho_s)=\frac{1}{\sqrt{\mathcal{A}(\rho_s)-\mathcal{A}(\rho)\mathcal{C}(\rho_s)/\mathcal{C}(\rho)}},
\end{equation}
where $z=1-\frac{\rho_s}{\rho}$. The function $R(z,\rho_s)$ is
regular for all values of $z$ and $\rho_s$, while $f(z,\rho_s)$
diverges as $z$ approaches to zero. Therefore, we can split the
integral~(\ref{Iz}) in two parts, the divergent part $I_D(\rho_s)$
and the regular one $I_R(\rho_s)$, which are given by,
\begin{equation}\label{Id}
I_D(\rho_s)=\int^1_0R(0,\rho_{ps})f_0(z,\rho_s)\,dz,
\end{equation}
\begin{equation}\label{Ir}
I_R(\rho_s)=\int^1_0\left[R(z,\rho_s)f(z,\rho_s)-R(0,\rho_{ps})f_0(z,\rho_s)\right]\,dz.
\end{equation}
Here, we expand the argument of the square root in $f(z,\rho_s)$ up
to the second order in $z$ \cite{SKKGBH},
\begin{equation}\label{f0}
f_0(z,\rho_s)=\frac{1}{\sqrt{p(\rho_s)z+q(\rho_s)z^2}},
\end{equation}
where
\begin{equation}\label{p}
p(\rho_s)=\frac{\rho_s}{\mathcal{C}(\rho_s)}\left[\mathcal{C}^\prime(\rho_s)\mathcal{A}(\rho_s)-\mathcal{C}(\rho_s)\mathcal{A}^\prime(\rho_s)\right],
\end{equation}
\begin{equation}\label{q}
{q}(\rho_s)=\frac{\rho_s^2}{2\mathcal{C}(\rho_s)}\left[2\mathcal{C}^\prime(\rho_s)\mathcal{C}(\rho_s)\mathcal{A}^\prime(\rho_s)
-2\mathcal{C}^\prime(\rho_s)^2\mathcal{A}(\rho_s)+\mathcal{A}(\rho_s)\mathcal{C}(\rho_s)\mathcal{C}^{\prime\prime}(\rho_s)-
\mathcal{C}^2(\rho_s)\mathcal{A}^{\prime\prime}(\rho_s)\right].
\end{equation}
For $\rho_s>\rho_{ps}$, $p(\rho_s)$ is nonzero and the leading order
of the divergence in $f_0$ is $z^{-1/2}$, which have a finite
result. As $\rho_s\longrightarrow\rho_{ps }$, $p(\rho_s)$ approaches
zero and divergence is of order $z^{-1}$, that makes the integral
divergent. Therefor, the deflection angle can be approximated in the
following form \cite{bozza1},
\begin{equation}\label{deflection}
\alpha=-\bar{a}\,log\left(\frac{u}{u_{sp}}-1\right)+\bar{b}+O(u-u_{sp}),
\end{equation}
where
\begin{eqnarray}\label{a}
&&\bar{a}=\frac{R(0,\rho_{ps})}{2\sqrt{q(\rho_{ps})}}\,,\nonumber\\
&&\bar{b}=-\pi+b_R+\bar{a}\,log\frac{\rho_{ps}^2\left[\mathcal{C}^{\prime\prime}(\rho_{ps})\mathcal{A}(\rho_{ps})-
\mathcal{C}(\rho_{ps})\mathcal{A}^{\prime\prime}(\rho_{ps})\right]}{u_{ps}\sqrt{\mathcal{A}^3(\rho_{ps})\mathcal{C}(\rho_{ps})}}\,,\nonumber\\
&&b_R=I_R(\rho_{ps}),\,\,\,\,\,\,u_{ps}=\sqrt{\frac{\mathcal{C}(\rho_{ps})}{\mathcal{A}(\rho_{ps})}}\,.
\end{eqnarray}
 By using ~(\ref{deflection})
and~(\ref{a}), we can investigate the properties of strong
gravitational lensing in the charged squashed Kaluza- Klein
G\"{o}del black hole. In this case, variations of the coefficients
$\bar{a}$ and $\bar{b}$, and the deflection angle $\alpha$ have been
plotted with respect to the extra dimension $\rho_0$, charge of the
black hole $\rho_q$, and G\"{o}del parameter $j$ in
 figures 4-6.\\
 As $j$ tends to zero, these quantities reduce to charged squashed Kaluza-klein black hole \cite{CSKKBH} and with $\rho_q=0$
 the squashed Kaluza-klein black hole recovers  \cite{SKKBH}.
 One can see that the deflection angle increases with extra
 dimension  and decreases with $\rho_q$.  By comparing these parameters with
 those in four-dimensional Schwarzschild and  Reissner-Nordstr\"{o}m
 black holes , we could extract information about the size of
 extra dimension as well as the charge of the black hole by using strong field
 gravitational lensing.
\begin{figure}
\centerline{\epsffile{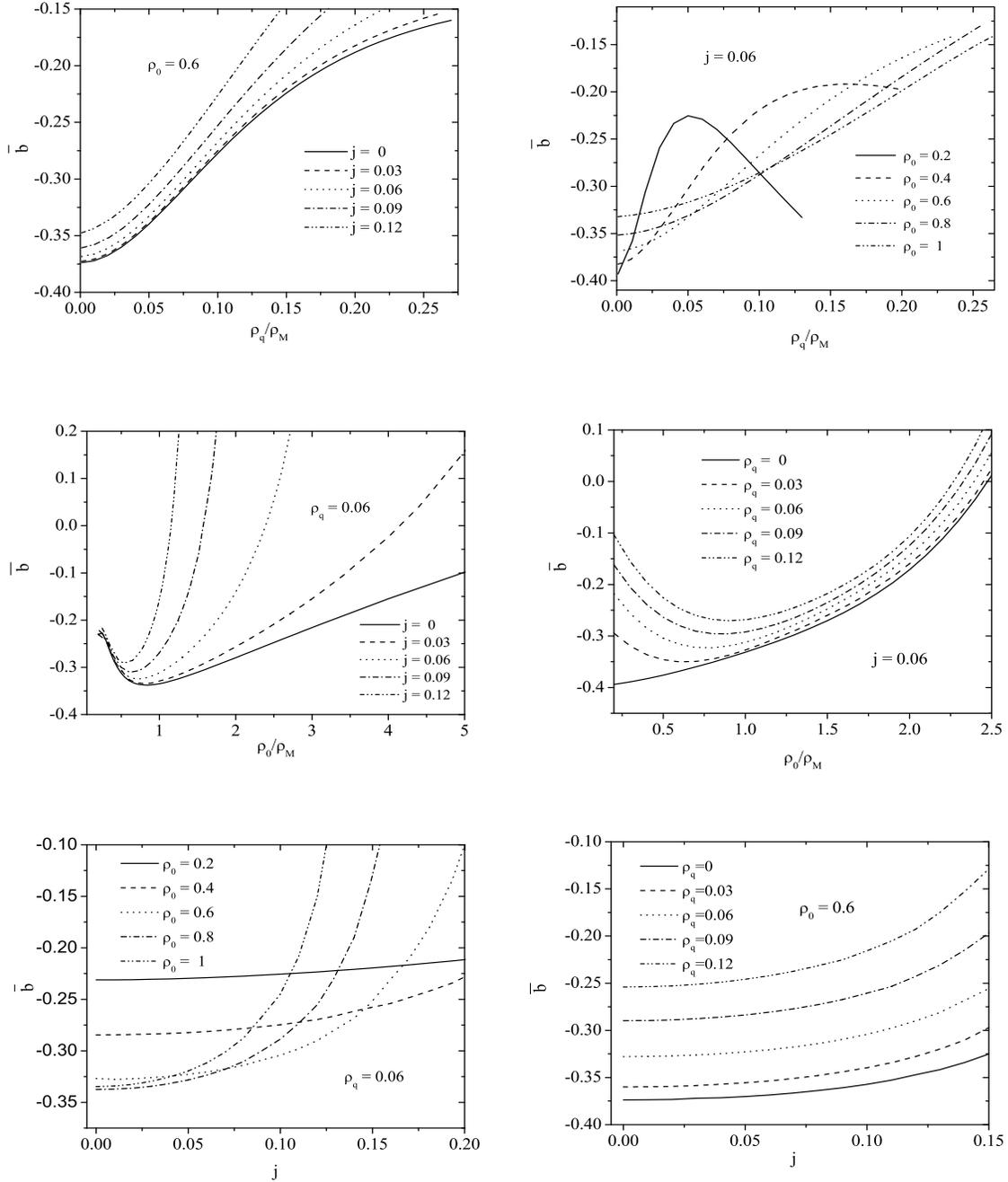}} \caption{The
 variation  of $\bar{b}$ with respect to $j$, $\rho_0$ and $\rho_q$.}
\end{figure}

\begin{figure}
\centerline{\epsffile{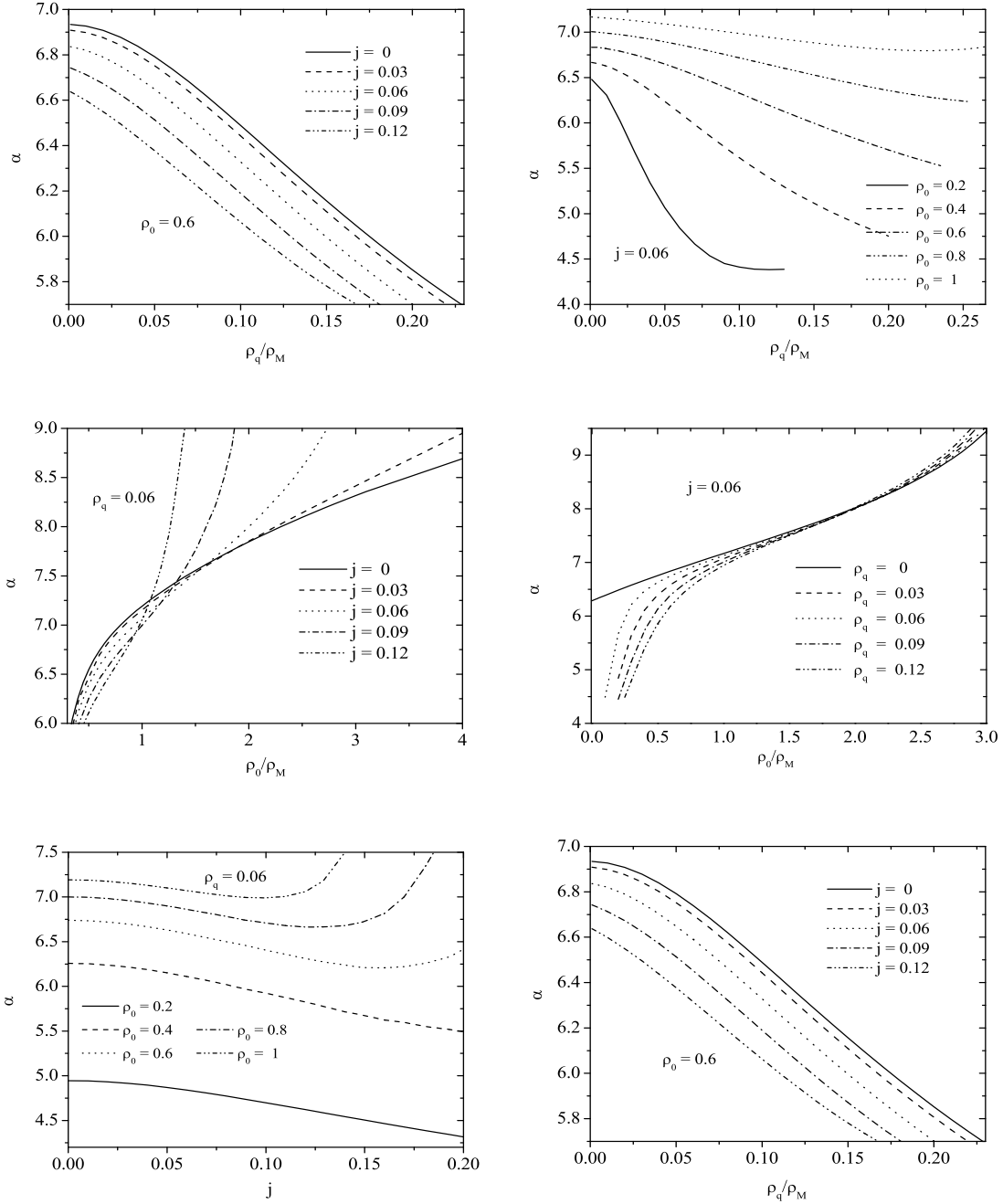}} \caption{Deflection angle as a
function of $j$, $\rho_0$ and $\rho_q$ at $x_s=x_{ps}+0.05$. Note
that $\alpha$ is given in Radian.}
\end{figure}

\begin{figure}
\centerline{\epsffile{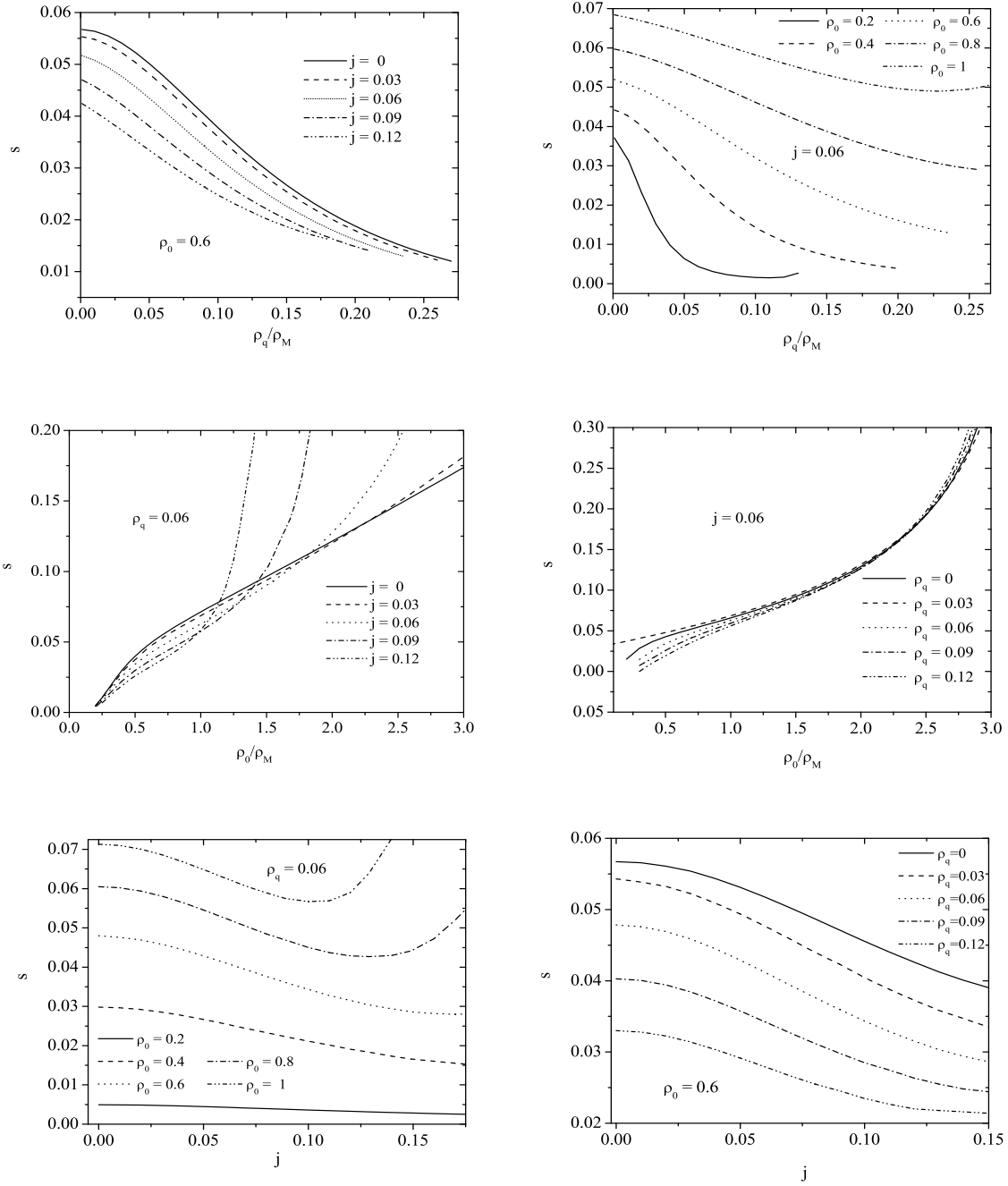}} \caption{The
 variation  of $s$ with respect to $j$, $\rho_0$ and $\rho_q$ . The angular separation is expressed in
$\mu$arcseconds. }
\end{figure}

\begin{table*}[s]
\begin{center}
\begin{tabular}{c | c | c c c | c c c |c c c}
  \hline
  \hline
    $\rho_q$      &    $\rho_0$  &     &      $\theta_{\infty}$    & &  &  $s$\    &  & &   $r_m$     \\
  \hline
    &        & $j=0$       &     $j=0.03$         &    $j=0.06$       & $j=0$       &     $j=0.03$         &    $j=0.06$     & $j=0$       &     $j=0.03$         &    $j=0.06$        \\
  \hline
    &  $0$   & 26.007  & 25.473 & 24.013 &0.0325 & 0.0319 &0.0300&6.8219& 6.8219 & 6.8219 \\
     &  $0.2$   &27.669  & 26.962    &  25.080 & 0.0339 & 0.0390& 0.0365 &6.6838 & 6.8212 &6.6736\\
  {}   & $0.4$    &29.214 & 28.327 & 25.987 & 0.0476 &0.0465 & 0.0434 &6.5678&6.5617&6.5440\\
  0    & $0.6$    &30.662 & 29.583 & 26.735 & 0.0556 & 0.0543 & 0.0507& 6.4681&6.4583& 6.4269\\
  {}   & $0.8$    &32.032 & 30.749 & 27.370 & 0.0639 & 0.0625 &  0.0587&6.3830&6.3660&6.3188\\
  {}   & $1$      &33.335 & 31.836 & 27.899 & 0.0726 & 0.0710  & 0.0672&6.3046&6.2835&6.2167\\
  \hline
  {}   &  $0.2$    & 29.389 & 28.121 & 25.005 & 0.0171 & 0.0164 & 0.0150&7.7688&7.7606&7.7378  \\
  {}   & $0.4$     &29.260 & 27.914 & 24/574 & 0.0411 & 0.0397& 0.0361&6.7556&6.7450&6.7142 \\
  0.03 & $0.6$     &30.594 & 29.093 & 25.339 & 0.0531 & 0.0512& 0.0477&6.5281&6.5136&6.4706  \\
  {}   & $0.8$     &31.953 & 30.258 & 26.000 & 0.0627 & 0.0607& 0.0557&6.4063&6.3872&6.3270\\
  {}   & $1$       &33.261 & 31.344 & 26.525 & 0.0719 & 0.0697 & 0.0647&6.3163&6.2907&6.2107 \\
  \hline
  {}   &  $0.2$   & 34.882 & 33.270 & 29.337 & 0.0048 & 0.0047 & 0.0042 & 9.3012 & 9.2919 & 9.2658   \\
  {}   & $0.4$    &  29.395 & 27.677 & 23.623 & 0.0292 & 0.0281 & 0.0251 & 7.1808 & 7.1680 & 7.1155  \\
  0.06 & $0.6$    & 30.412 & 28.527 & 24.031 & 0.0469 & 0.0450 & 0.0404 & 6.6806 & 6.6598 & 6.5995 \\
  {}   & $0.8$    & 31.738 & 29.648 & 24.632 & 0.0593 & 0.0575 &0.0515 & 6.4727& 6.4470 & 6.3704  \\
  {}   & $1$      & 33.052 & 30.736 & 25.132 & 0.0699 & 0.0674 & 0.0616 & 6.3480 & 6.3159 & 6.2190  \\
  \hline
  {}   & $0.2$ & 48.924 & 46.807 & 41.546 & 0.0021 & 0.0020 & 0.0018 & 10.476 & 10.470 & 10.452   \\
  {}   & $0.4$ &  29.635 & 27.645 & 23.073 & 0.0191 & 0.0183 & 0.0163 & 7.6851 & 7.6624 & 7.6058   \\
  0.09 & $0.6$ &  30.144 & 27.932 & 22.858 & 0.0403 & 0.0338 & 0.0337 & 6.8883 & 6.8601 & 6.7795 \\
  {}   & $0.8$ & 31.408 & 28.961 & 23.318 & 0.0548 & 0.0523 & 0.0468 & 6.5684 & 6.5351 & 6.4372 \\
  {}   & $1$ & 32.732 & 30.027 & 23.761 & 0.0672 & 0.0643 & 0.0582 & 6.3948 & 6.3555 & 6.2362    \\
  \hline
  {}   & $0.2$ & 111.910 & 107.525  & 96.386   & 0.0019 & 0.0018  & 0.0017  & 11.362 & 11.360 & 11.353  \\
  {}   & $0.4$ & 30.003 & 27.821  & 22.893   & 0.0124 & 0.0118  & 0.0105 &  8.1843 & 8.1580&  8.0848   \\
  0.12 & $0.6$ &   29.809 & 27.331   & 21.824 & 0.0324 & 0.0308  & 0.0274 &7.1226
  & 7.0867 & 6.9846 \\
  {}   & $0.8$ &   30.984 & 28.221  & 22.075  & 0.0497 & 0.0473   & 0.0422    &6.6823
& 6.6418 & 6.5175\\
  {}   & $1$ &    32.311 & 29.255  &  22.433   & 0.0639 & 0.0610 & 0.0550  & 6.4516&
  6.4031 & 6.2563
  \\
  \hline
\end{tabular}
\caption {Numerical estimations for the coefficients and observables
of strong gravitational lensing by considering the supermmasive
object of galactic center be a charged squashed Kaluza-Klein
G\"{o}del black hole. (Not that the numerical values for
$\theta_{\infty}$ and $s$ are of order microarcsec)   }
\end{center}
\end{table*}
\section{Observables  estimation}
In the previous section,  we investigated the strong gravitational
lensing by using a simple and reliable logarithmic formula for
deflection angle, which was obtain by Bozza et al. Now, by using
relations between the parameters of the strong gravitational lensing
and observables, estimat the position and magnification of the
relativistic images.  By comparing these observables with the data
from the astronomical observation, we could detect properties of an
massive object.  We suppose that the spacetime of the supermassive
object at the galaxy center of Milky Way can be considered as a
charged squashed Kaluza-Klein G\"{o}del black hole, then we can
estimate the numerical values for observables.\\
We can write the lens equation in strong gravitational lensing, as
the source, lens, and observer are highly aligned as follows
\cite{bozza2},
\begin{equation}\label{lensEQ}
\beta=\theta-\frac{D_{LS}}{D_{OS}}\Delta\alpha_n,
\end{equation}\\
where $D_{LS}$ is the distance between the lens and source. $D_{OS}$
is the distance between the observer and the source so that,
$D_{OS}=D_{LS}+D_{OL}$. $\beta$ and $\theta$ are the angular
position of the source and the image with respect to lens,
respectively. $\Delta\alpha_n=\alpha-2n\pi$ is the offset of
deflection angle
with integer $n$  which indicates the $n$-th image.\\
The $n$-th image position $\theta_n$ and the $n$-th image
magnification $\mu_n$ can be approximated as follows
\cite{bozza2,bozza1},
\begin{figure}
\centerline{\epsffile{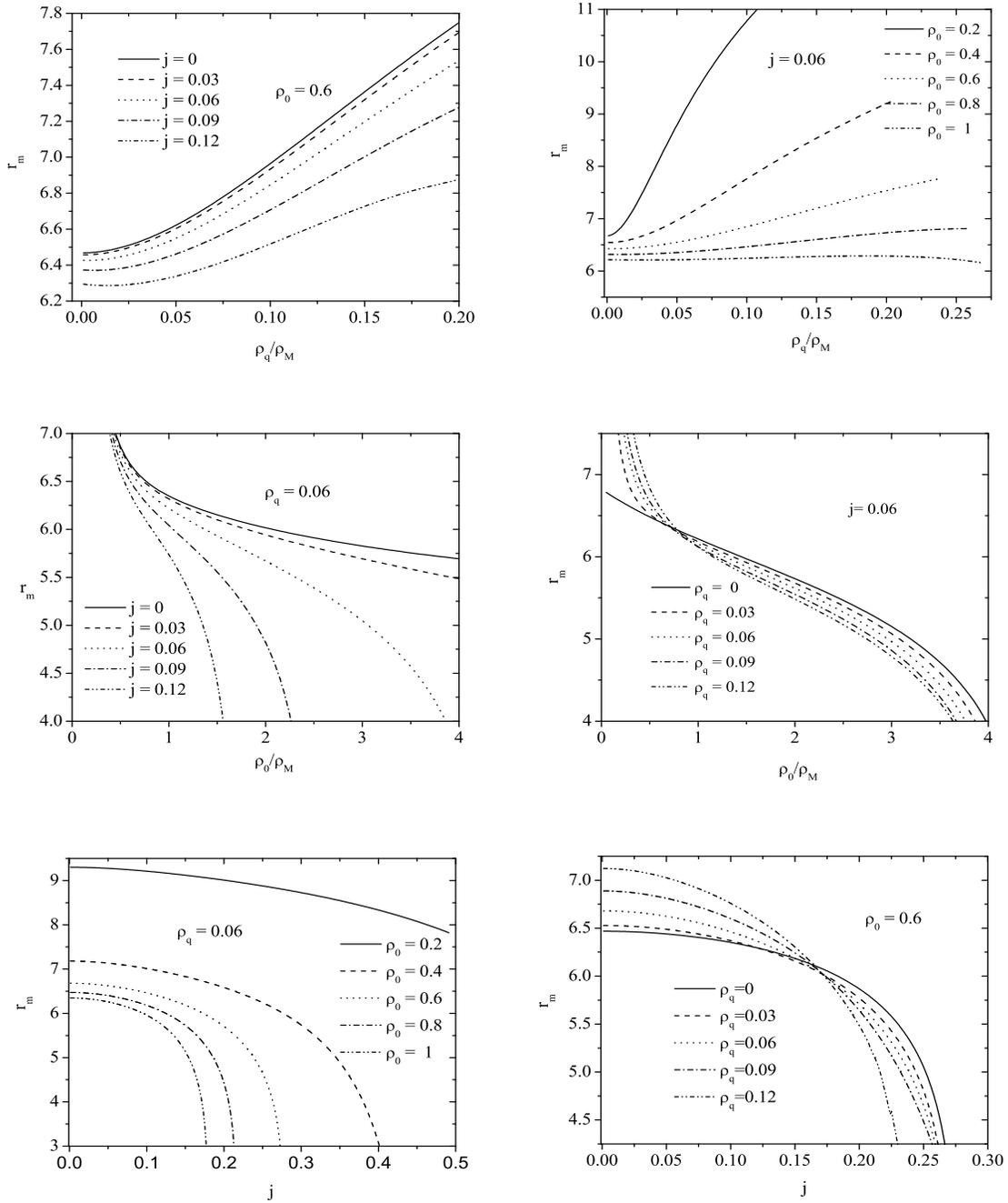}} \caption{The variation of  $r_m$ with
$j$, $\rho_0$ and $\rho_q$.}
\end{figure}
\begin{figure}
\centerline{\epsffile{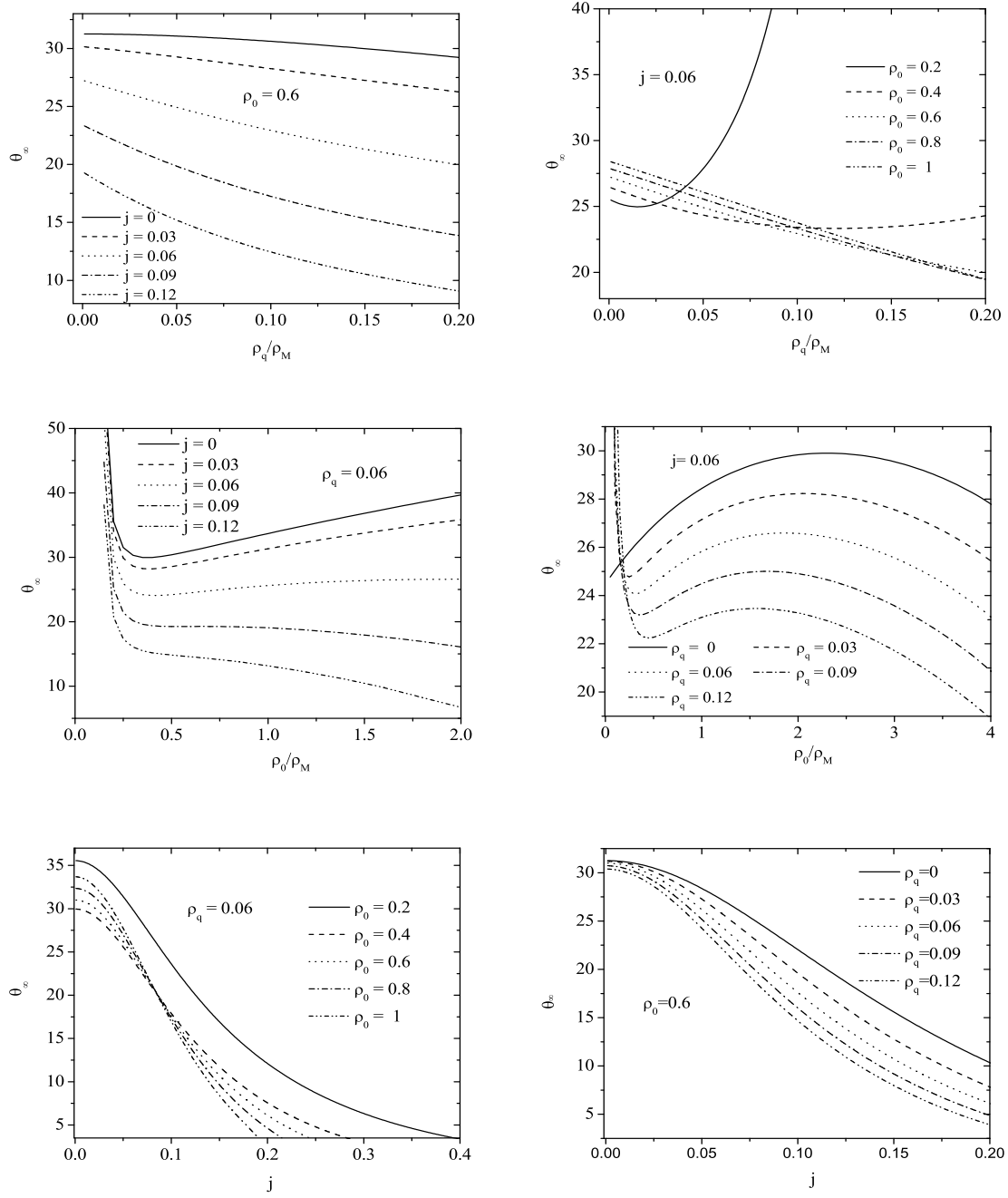}} \caption{The variation of angular
position $\theta_{\infty}$ with respect to $j$, $\rho_0$ and
$\rho_q$ that is given in $\mu$arcseconds.}
\end{figure}
\begin{equation}\label{theta}
\theta_n=\theta^0_n+\frac{u_{ps}(\beta-\theta_n^0)e^{\frac{\bar{b}-2n\pi}{\bar{a}}}D_{OS}}{\bar{a}
D_{LS}D_{OL}},
\end{equation}
\begin{equation}\label{magnification}
\mu_n=\frac{u_{ps}^2(1+e^{\frac{\bar{b}-2n\pi}{\bar{a}}})e^{\frac{\bar{b}-2n\pi}{\bar{a}}}D_{OS}}{\bar{a}\beta
D_{LS}D_{OL}^2}.
\end{equation}
$\theta_n^0$ is the angular position of $\alpha=2n\pi$. In the limit
$n\longrightarrow\infty$, the relation between the minimum of impact
parameter $u_{ps}$ and asymptotic position of a set of images
$\theta_{\infty}$ can be expressed by
$u_{ps}=D_{OL}\theta_{\infty}$. In order to obtain the coefficients
$\bar{a}$ and $\bar{b}$, in the simplest case, we separate the
outermost image $\theta_1$ and all the remaining ones which are
packed together at $\theta_{\infty}$, as done in Refs
\cite{bozza2,bozza1}. Thus $s=\theta_1-\theta_{\infty}$ is
considered as the angular separation between the first image and
other ones and the ratio of the flux of them is given by,
\begin{equation}\label{R}
\mathcal{R}=\frac{\mu_1}{\sum_{n=2}^{\infty}\mu_n}.
\end{equation}
We can simplify the observables and rewrite them in the following
form \cite{bozza2,bozza1},
\begin{eqnarray}\label{sR}
&&s=\theta_{\infty}e^{\frac{\bar{b}}{\bar{a}}-\frac{2\pi}{\bar{a}}},\nonumber\\
&&\mathcal{R}=e^{\frac{2\pi}{\bar{a}}}.
\end{eqnarray}
Thus, by measuring the $s$, $\mathcal{R}$ and $\theta_{\infty}$, one
can obtain the  values of the coefficients $\bar{a}$, $\bar{b}$ and
$u_{sp}$. If we compare these  values by those obtained  in the
previous section, we could detect the size of the extra dimension,
charge of black hole and rotation of universe.  Another  observable
for gravitational lensing is relative magnification of the outermost
relativistic image with the other ones. This observable is shown by
$r_m$ which is related to $\mathcal{R} $ by,
\begin{eqnarray}\label{rm}
r_m=2.5\, \log\mathcal{R}.
\end{eqnarray}
Using $\theta_{\infty}=\frac{u_{sp}}{D_{OL}}$ and
equations~(\ref{a}),~(\ref{sR}) and~(\ref{rm}) we can estimate the
values of the observable in the strong field gravitational lensing.
The variation of  the observables $\theta_{\infty}$, $s$ and $r_m$
are plotted in figures 7-9. Note that the mass of the central object
of our galaxy is estimated to be $4.31\times 10^6 M_\odot$ and  the
distance between the sun and the
center of galaxy is $D_{OL}=8.5\,kpc$ \cite{Gillessen}.\\
For  different $\rho_0$, $\rho_q$ and $j$, the numerical values for
the observables  are listed in Table 1. One can see that our results
reduce to those in the four-dimensional Schwarzschild black hole as
$\rho_0\longrightarrow0$. Also our results are in agreement with the
results of Ref. \cite{SKKGBH} in the limit $\rho_q\longrightarrow0$
and  in the limit $j\longrightarrow0$, the results of Ref.
\cite{CSKKBH} are recovered.
\section{Summary}
The light rays can be deviated from the straight way in the
gravitational field as predicted by General Relativity. This
deflection of light rays is known as gravitational lensing. In the
strong field limit, the deflection anglethe of the light rays which
pass very close to the black hole, becomes so large  that, it winds
several times around the black hole before appearing at the
observer. Therefore the observer would detect two infinite set of
faint relativistic images produced on each side of the black hole.
On the other hand the extra dimension is one of the important
predictions in the string theory which is believed to be a promising
candidate for the unified theory. Also it is reasonable to consider
a rotative universe with global rotation. Hence the five-dimensional
Einstein-Maxwell theory with a Chern- Simons term in string theory
predicted five-dimensional charged black holes in the G\"{o}del
background. In our study, we considered the charged squashed
Kaluza-Klein G\"{o}del black hole spacetime and investigated the
strong gravitational lensing by this metric. We  obtained
theoretically the deflection angle and other parameters of strong
gravitational lensing . Finally, we suppose that the supermassive
black hole at the galaxy center of Milky Way can be considered by
this spacetime and we estimated numerically the values of
observables that are realated to the lensing parameters. Theses
observable parameters are $\theta_{\infty}$, $s$ and $R$, where
$\theta_{\infty}$ is the position of relativistic images, $s$
angular separation between the first image $\theta_1$ and other ones
$\theta_{\infty}$ and $R$ is the ratio of the flux from the first
image  and those from all the other images.  Our results are
presented in figures 1-9 and Table 1. By comparatione observable
parameters with observational data measured by the astronomical
instruments in the future, we can discuss  the properties of the
massive object in the center of our galaxy.

\subsection*{Acknowledgments}
 H. Vaez would like to thank Afshin. Ghari for helpful comments.

\end{document}